\documentclass[12pt,preprint]{aastex}

\begin{document}

\title{Photometric and period investigation of the late F-type overcontact binary II UMa}

\author{X. Zhou\altaffilmark{1,2,3}, S.-B. Qian\altaffilmark{1,2,3}, J. Zhang\altaffilmark{1,2}, B. Zhang\altaffilmark{1,2,3}, J. Kreiner\altaffilmark{4}}

\singlespace

\altaffiltext{1}{Yunnan Observatories, Chinese Academy of Sciences (CAS), P. O. Box 110, 650216 Kunming, China; zhouxiaophy@ynao.ac.cn}
\altaffiltext{2}{Key Laboratory of the Structure and Evolution of Celestial Objects, Chinese Academy of Sciences, P. O. Box 110, 650216 Kunming, China}
\altaffiltext{3}{University of Chinese Academy of Sciences, Yuquan Road 19\#, Sijingshang Block, 100049 Beijing, China}
\altaffiltext{4}{Mt. Suhora Astronomical Observatory, Pedagogical University of Cracow, Poland}
\begin{abstract}
II UMa is a late F-type (F5) contact binary with a close-in tertiary and a distant visual companion. According to the four-color ($B$ $V$ $R_c$ $I_c$) light curves' solutions of II UMa, it is a high fill-out (f=$86.6\,\%$) and low mass ratio ($q = 0.172$) contact binary system, which indicate that it is at the late evolutionary stage of late-type tidal-locked binary stars. The mass of the primary star and secondary one are calculated to be $M_1 = 1.99M_\odot$, $M_2 = 0.34M_\odot$. The primary star has evolved from ZAMS, but it is still before TAMS, and the secondary star is even more evolved. Considering the mass ratio ($M_3/M_1 = 0.67$) obtained by spectroscopic observations, the mass of the close-in tertiary is estimated to be $M_3 = 1.34M_\odot$. The period variations of the binary system is investigated for the first time. According to the observed-calculated ($O$-$C$) curve analysis, a continuous period increase at a rate of $dP/dt=4.88\times{10^{-7}}day\cdot year^{-1}$ is determined. It may be just a part of a cyclic period change, or the combinational period change of a parabolic variation and a cyclic one. More times of minimum light are needed to confirm this. The presence of the tertiary component may play an important role in the formation and evolution of this binary system by drawing angular momentum from the central system during the pre-contact stage.

\end{abstract}

\keywords{binaries : close --
          binaries : eclipsing --
          stars: evolution --
          stars: individuals (II UMa)}

\section{Introduction}
W UMa type binaries are cool short-period (usually less than 1 day) binary systems with both components filling their critical Roche lobes and sharing a common convective envelope during their main sequence (MS) evolutionary stage. The more massive primary component is a main sequence star whereas the secondary is oversized compared to its expected MS radius. The formation and evolution of W UMa type binary systems are still unsolved problems in astrophysics. The most popular evolutionary scenario is that they are formed from initially detached systems via angular momentum loss (AML) by means of magnetic stellar wind \citep{1982A&A...109...17V,1989AJ.....97..431E}. Model calculations suggest that these binary stars will ultimately coalesce into single stars which may be progenitors of the poorly understood blue stragglers and FK Com-type stars \citep{2006AcA....56..199S,2011AcA....61..139S}.

II UMa (BD+55 1540, HIP 61237, $V$ = $8^{m}.48$) is a component of the visual binary ADS 8954, with the separation of $0''.87$ and difference in brightness of 1.64 mag. The photometric variability of the star was discovered by the $Hipparcos$ satellite (ESA, 1997). Radial velocity curves of both components in II UMa were obtained by \citet{2002AJ....124.1738R} and gave the results: $q = 0.172\pm0.004$, $V_\gamma = -8.02\pm1.10$ km s$^{-1}$, $(M_1+M_2)sin^3i = 2.180\pm0.080M_\odot$. They pointed out that II UMa was a A-subtype W UMa binary system with a spectral type of F5III. Later, detailed investigation by \citet{2006AJ....132..650D} determined that it was a triple system with a solar-type close-in tertiary component. The temperature of the tertiary component is about $6100K$ and the mass ratio is $M_3/M_1 = 0.67$. The first photometric solutions of II UMa were published by \citet{2007ASPC..362...82O}. Then, \citet{2015NewA...34..271Y} also obtained the photometric parameters of this binary system. In the present paper, four-color light curves (LCs) of II UMa are analyzed and its formation and evolutionary scenario are discussed. The period variations of the binary system is investigated for the first time, which may reveal the dynamics interaction between the two components.

\section{Observations}
The four-color ($B$ $V$ $R_c$ $I_c$) light curves of II UMa were carried out in five continuous nights on February 1, February 2, February 3, February 4 and  February 5, 2012 with an Andor DV436 2K CCD camera attached to the 60cm reflecting telescope at Yunnan Observatories (YNOs). The coordinates of the variable star, the comparison star and the check star were listed in Table \ref{Coordinates2}. The integration time were 60s for $B$ band, 30s for $V$ band, 15s for $R_c$ band, and 10s for $I_c$ band, respectively. The light curves of those observations were displayed in Fig. 1. During the observations, the broadband Johnson-Cousins $B$ $V$ $R_c$ $I_c$ filters were used. PHOT (measured magnitudes for a list of stars) of the aperture photometry package in the IRAF \footnote {The Image Reduction and Analysis Facility is hosted by the National Optical Astronomy Observatories in Tucson, Arizona at URL iraf.noao.edu.} was used to reduce the observed images.

\begin{table}[!h]
\begin{center}
\caption{Coordinates of II UMa, the comparison and the check stars.}\label{Coordinates2}
\begin{small}
\begin{tabular}{ccccc}\hline\hline
Targets          &   name               & $\alpha_{2000}$           &  $\delta_{2000}$         &  $V_{mag}$     \\ \hline
Variable         &   II  UMa            &$12^{h}32^{m}54^{s}.8$     & $+54^\circ47'42''.9$     &  $8.26$           \\
The comparison   &   GSC 03841-00375    &$12^{h}32^{m}16^{s}.6$     & $+54^\circ47'57''.5$     &  $10.05$     \\
The check        &   GSC 03841-00259	&$12^{h}32^{m}19^{s}.2$     & $+54^\circ53'21''.7$     &  $11.15$     \\
\hline\hline
\end{tabular}
\end{small}
\end{center}
\end{table}

\begin{figure}[!h]
\begin{center}
\includegraphics[width=13cm]{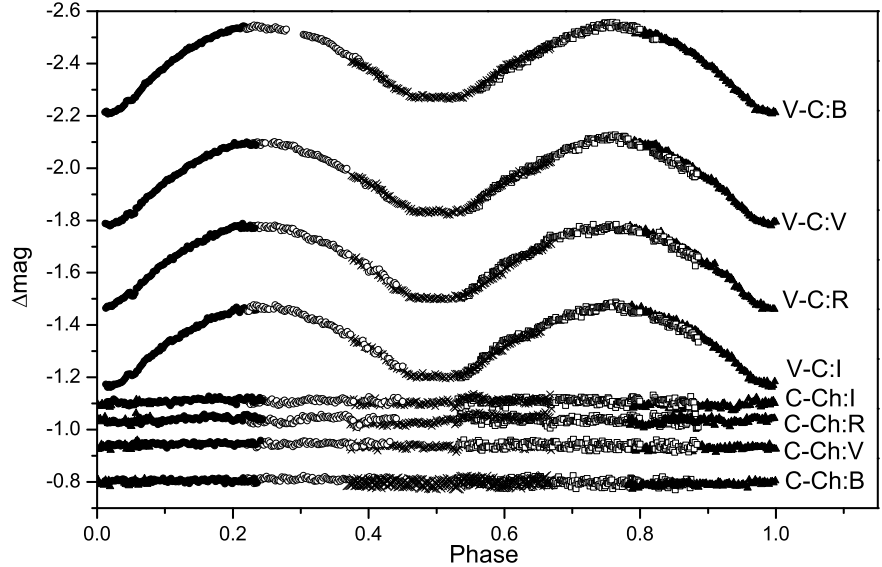}
\caption{CCD photometric light curves in $B$ $V$ $R_c$ and $I_c$ bands. The magnitude difference between the comparison and the check stars are presented. The standard deviations of the comparison-check observations are 0.010 mag for $B$ band, 0.010 mag for $V$ band, 0.012 mag for $R_c$ band and 0.011 mag for $I_c$ band. Solid circles, open circles, crosses, squares and triangles correspond to the data observed on February 1, February 2, February 3, February 4 and  February 5, 2012, respectively.}
\end{center}
\end{figure}

Times of minimum light of II UMa were also observed and determined, which were listed in Table \ref{Newminimum}.

\begin{table}[!h]
\begin{center}
\caption{New CCD times of light minimum for II UMa.}\label{Newminimum}
\begin{tabular}{cccccc}\hline
    JD (Hel.)     &  Error (days)  & Min. &   Filter      & Method  &Telescopes\\\hline
  2455961.3376    & $\pm0.0005$    &   II &   $BVR_cI_c$  &  CCD    &    60cm   \\
  2455963.3963    & $\pm0.0005$    &   I  &   $BVR_cI_c$  &  CCD    &    60cm   \\
  2456030.2416    & $\pm0.0002$    &   I  &   $N$         &  CCD    &    60cm   \\
  2456265.4368    & $\pm0.0011$    &   I  &   $R_cI_cN$   &  CCD    &    60cm   \\
  2456399.1215    & $\pm0.0008$    &   I  &   $VR_cI_cN$  &  CCD    &    60cm   \\
  2456404.0720    & $\pm0.0007$    &   I  &   $VR_cI_cN$  &  CCD    &    60cm   \\
  2456744.0597    & $\pm0.0007$    &   I  &   $VR_cI_c$   &  CCD    &    1m   \\
\hline
\end{tabular}
\end{center}
\textbf
{\footnotesize Notes.} \footnotesize 60cm and 1m denote to the 60cm and 1m reflecting telescope in Yunnan Observatories.
\end{table}

\section{Orbital Period Investigation}

The study of orbital period change is a very important part for contact binary stars. However, the period change investigation of II UMa has been neglected since it was discovered. During the present work, all available times of minimum light are collected. Minimum times with the same epoch have been averaged, and only the mean values are listed in Table \ref{Minimum}.

The minimum times of II UMa are listed in Table \ref{Minimum}. Using the following linear ephemeris,
\begin{equation}
Min.I(HJD) = 2456030.2416+0^{d}.82522\times{E},\label{linear ephemeris}
\end{equation}
the $O - C$ values are calculated and listed in the fourth column of Table \ref{Minimum} and plotted in the upper panel of Fig. 2. Based on the least-square method, the new ephemeris is
\begin{equation}
\begin{array}{lll}
Min. I = 2456030.2428(\pm0.0001)+0.82522813(\pm0.00000006)\times{E}
         \\+0.552(\pm0.007)\times{10^{-9}}\times{E^{2}}
\end{array}
\end{equation}
With the quadratic term included in this ephemeris, a continuous period increase, at a rate of $dP/dt=4.88\times{10^{-7}}day\cdot year^{-1}$ is determined. The residuals from Equation (2) are displayed in the lower panel of Fig. 2.

\begin{table}[!h]
\caption{$(O-C)$ values of minimum light for II UMa.}\label{Minimum}
\begin{center}
\small
\begin{tabular}{cclllcc}\hline\hline
JD (Hel.)      &  Min &   Epoch     & $(O-C)$      &   Error       & Method       &  Reference       \\
(2400000+)     &      &             &              &               &              &\\\hline
48371.7587     & II   &  -9280.5	& 	-0.0287    &    0.0014     &    CCD       &  1    \\
48372.1747     & I    &  -9280      &   -0.0253    &    0.0016     &    CCD       &  1    \\
51221.6579     & I    &  -5827	    & 	-0.0268    &    0.0044     &    CCD       &  2    \\
52649.7056     & II   &  -4096.5    &   -0.0223    &    0.0007     &    CCD       &  3    \\
52723.5654     & I    &  -4007	    & 	-0.0197    &    0.0004     &    CCD       &  4    \\
53064.3778     & I    &  -3594      &   -0.0231    &    0.0002     &    CCD       &  5    \\
53081.7130     & I    &  -3573	    &   -0.0176    &    0.0020     &    CCD       &  6    \\
53761.2818     & II   &  -2749.5    &   -0.0174    &               &    CCD       &  7    \\
53809.1465     & II   &  -2691.5 	&   -0.0155    &               &    CCD       &  7    \\
54528.3212     & I    &  -1820	    &   -0.0200    &               &    CCD       &  8    \\
55231.8365     & II   &  -967.5	    &   -0.0048    &    0.0001     &    CCD       &  9    \\
55280.5232     & II   &  -908.5	    &   -0.0060    &    0.0003     &    CCD       &  10    \\
55961.3376     & II   &  -83.5	    &   0.0018     &    0.0005     &    CCD       &  11    \\
55963.3963     & I    &  -81     	&   -0.0025    &    0.0005     &    CCD       &  11    \\
56030.2416     & I    &  0      	&   0          &    0.0002     &    CCD       &  11    \\
56057.4753     & I    &  33	        &   0.0014     &    0.0004     &    CCD       &  12    \\
56265.4368     & I    &  285     	&   0.0075     &    0.0011     &    CCD       &  11    \\
56319.4864     & II   &  350.5     	& 	0.0052     &    0.0006     &    CCD       &  12    \\
56399.1215     & I    &  447    	& 	0.0066     &    0.0008     &    CCD       &  11    \\
56404.0720     & I    &  453    	& 	0.0057     &    0.0007     &    CCD       &  11    \\
56713.5344     & I    &  828	    &   0.0106     &    0.0007     &    CCD       &  12    \\
56725.4998     & II   &  842.5	    &   0.0103     &    0.0002     &    CCD       &  12    \\
56744.0597     & I    &  865	    &   0.0027     &    0.0007     &    CCD       &  11    \\\hline
\end{tabular}
\end{center}
\textbf
{\footnotesize Reference:} \footnotesize (1) private provision; (2) \citet{2002AJ....124.1738R}; (3) \citet{2005IBVS.5623....1D}; (4) \citet{2005IBVS.5606....1P}; (5) \citet{2005IBVS.5592....1K};
(6) \citet{2005IBVS.5602....1N}; (7) \citet{2007VSB.45....1D}; (8) \citet{2009VSB.48....1D}; (9) \citet{2011IBVS.5974....1D}; (10) \citet{2011OEJV..137....1B}; (11) present work; (12) \citet{2014IBVS.6114....1Z}
\end{table}

\begin{figure}[!h]
\begin{center}
\includegraphics[width=13cm]{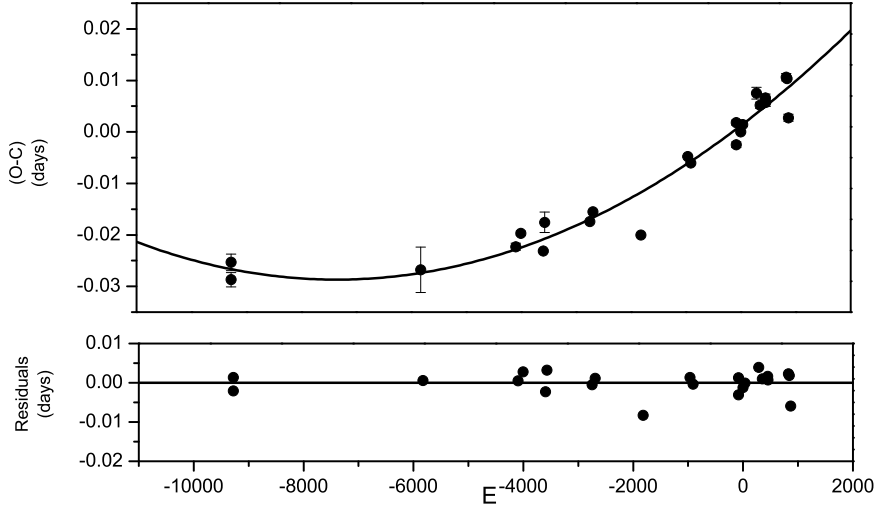}
\caption{The $(O-C)$ values of II UMa from the linear ephemeris of Equation (1) is presented in the upper panel. The solid line in the panel refers to upward parabolic variation, which reveals a continuous increase in the orbital period. After the upward parabolic change were removed, the residuals are plotted in the lower panel.}
\end{center}
\end{figure}

\section{Photometric Solutions}
II UMa is an EW-type contact binary. The W-D program of 2013 version \citep{Wilson1971,Wilson1979,Wilson1990,Van2007,Wilson2008,Wilson2010,Wilson2012} are used in modeling the light curves. The number of observational data points used in W-D program are 566 in $B$ band, 599 in $V$ band, 613 in $R_c$ band, 633 in $I_c$ band, respectively. The phases are calculated with the following linear ephemeris:
\begin{equation}
Min.I(HJD) = 2456030.2416+0^{d}.82522\times{E}.\label{linear ephemeris}
\end{equation}
According to its spectral type of F5, the effective temperature of star 1 is assumed to be $T_1=6550$K. The bolometric albedo $A_1=A_2=0.5$ \citep{1969AcA....19..245R} and the values of the gravity-darkening coefficients $g_1=g_2=0.32$ \citep{1967ZA.....65...89L} are used. The bolometric and passband-specific limb-darkening coefficients are chosen from \citet{1993AJ....106.2096V}'s table. The adjustable parameters are: the orbital inclination $i$; the mean temperature of star 2 ($T_{2}$); the monochromatic luminosity of star 1 ($L_{1B}$, $L_{1V}$, $L_{1R}$ and $L_{1I}$); the dimensionless potential of star 1 ($\Omega_{1}=\Omega_{2}$ in mode 3 for overcontact configuration) and the third light $l_3$. The final photometric solutions are listed in Table 4 and the theoretical light curves are displayed in Fig. 3. The contact configuration of II UMa is displayed in Fig. 4.

\begin{table}[!h]
\begin{center}
\caption{Photometric solutions of II UMa}\label{phsolutions}
\small
\begin{tabular}{lllllllll}
\hline\hline
Parameters                            &  Solution A                   &  Solution B                   &  Solution C      \\\hline
$T_{1}(K)   $                         &  6550(fixed)                  &  6680(fixed)                  &  6412(fixed)   \\
$g_{1}$                               &  0.32(fixed)                  &  0.32(fixed)                  &  0.32(fixed)\\
$g_{2}$                               &  0.32(fixed)                  &  0.32(fixed)                  &  0.32(fixed) \\
$A_{1}$                               &  0.50(fixed)                  &  0.50(fixed)                  &  0.50(fixed)  \\
$A_{2}$                               &  0.50(fixed)                  &  0.50(fixed)                  &  0.50(fixed)  \\
q ($M_2/M_1$ )                        &  0.172(fixed)                 &  0.172(fixed)                 &  0.172(fixed)  \\
$i(^{\circ})$                         &  77.8($\pm0.3$)               &  77.8($\pm0.3$)               &  77.7($\pm0.3$)  \\
$\Omega_{in}$                         &  2.1615                       &  2.1615                       &  2.1615    \\
$\Omega_{out}$                        &  2.0511                       &  2.0511                       &  2.0511  \\
$\Omega_{1}=\Omega_{2}$               &  2.0659($\pm0.0047$)          &  2.0685($\pm0.0057$)          &  2.0656($\pm0.0048$)\\
$T_{2}(K)$                            &  6554($\pm7)$                 &  6684($\pm8$)                 &  6418($\pm7$)   \\
$\Delta T(K)$                         &  4                            &  4                            &  6  \\
$T_{2}/T_{1}$                         &  1.001($\pm0.001$)            &  1.001($\pm0.001$)            &  1.001($\pm0.001$)  \\
$L_{1}/(L_{1}+L_{2}$) ($B$)           &  0.8084($\pm0.0008$)          &  0.8086($\pm0.0007$)          &  0.8082($\pm0.0008$)   \\
$L_{1}/(L_{1}+L_{2}$) ($V$)           &  0.8091($\pm0.0010$)          &  0.8092($\pm0.0010$)          &  0.8089($\pm0.0010$)    \\
$L_{1}/(L_{1}+L_{2}$) ($R_c$)         &  0.8094($\pm0.0015$)          &  0.8095($\pm0.0015$)          &  0.8092($\pm0.0015$)   \\
$L_{1}/(L_{1}+L_{2}$) ($I_c$)         &  0.8097($\pm0.0030$)          &  0.8097($\pm0.0030$)          &  0.8095($\pm0.0030$)    \\
$L_{3}/(L_{1}+L_{2}+L_{3}$) ($B$)     &  0.2992($\pm0.0023$)          &  0.2954($\pm0.0023$)          &  0.3013($\pm0.0023$)          \\
$L_{3}/(L_{1}+L_{2}+L_{3}$) ($V$)     &  0.2754($\pm0.0032$)          &  0.2728($\pm0.0032$)          &  0.2762($\pm0.0032$)          \\
$L_{3}/(L_{1}+L_{2}+L_{3}$) ($R_c$)   &  0.2674($\pm0.0051$)          &  0.2648($\pm0.0051$)          &  0.2684($\pm0.0050$)          \\
$L_{3}/(L_{1}+L_{2}+L_{3}$) ($I_c$)   &  0.2682($\pm0.0098$)          &  0.2658($\pm0.0098$)          &  0.2693($\pm0.0097$)          \\
$r_{1}(pole)$                         &  0.5209($\pm0.0004$)          &  0.5241($\pm0.0008$)          &  0.5219($\pm0.0001$)    \\
$r_{1}(side)$                         &  0.5790($\pm0.0006$)          &  0.5842($\pm0.0014$)          &  0.5807($\pm0.0002$)     \\
$r_{1}(back)$                         &  0.6083($\pm0.0008$)          &  0.6148($\pm0.0017$)          &  0.6104($\pm0.0002$)    \\
$r_{2}(pole)$                         &  0.2499($\pm0.0005$)          &  0.2512($\pm0.0011$)          &  0.2522($\pm0.0002$)    \\
$r_{2}(side)$                         &  0.2644($\pm0.0006$)          &  0.2660($\pm0.0013$)          &  0.2672($\pm0.0002$)    \\
$r_{2}(back)$                         &  0.3424($\pm0.0024$)          &  0.3490($\pm0.0057$)          &  0.3546($\pm0.0010$)   \\
$f$                                   &  $86.6\,\%$($\pm$4.2\,\%$$)   &  $84.2\,\%$($\pm$5.2\,\%$$)   &  $86.9\,\%$($\pm$4.3\,\%$$)   \\
$\Sigma{\omega(O-C)^2}$               &  0.036625                     &  0.036671                     &  0.036514   \\
\hline
\end{tabular}
\end{center}
\textbf
{\footnotesize Notes.} \footnotesize The errors listed in Table 4 are internal errors resulting from the application of the WD code to the supplied data.
\end{table}

\begin{figure}[!h]
\begin{center}
\includegraphics[width=14cm]{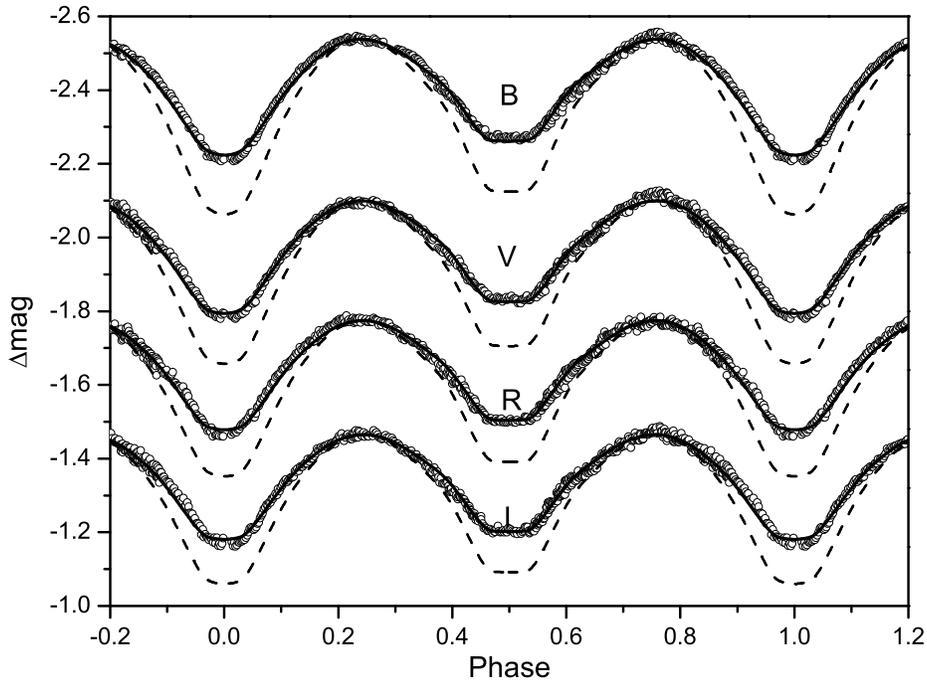}
\caption{Observed (open circles) and theoretical (solid lines) light curves in the $B V R_c$ and $I_c$ bands for II UMa. The standard deviation of the fitting residuals are 0.011 mag for $B$ band, 0.011 mag for $V$ band, 0.010 mag for $R_c$ band and 0.010 mag for $I_c$ band, respectively. The dash lines represent theoretical light curves without the third light.}
\end{center}
\end{figure}

\begin{figure}[!h]
\begin{center}
\includegraphics[width=14cm]{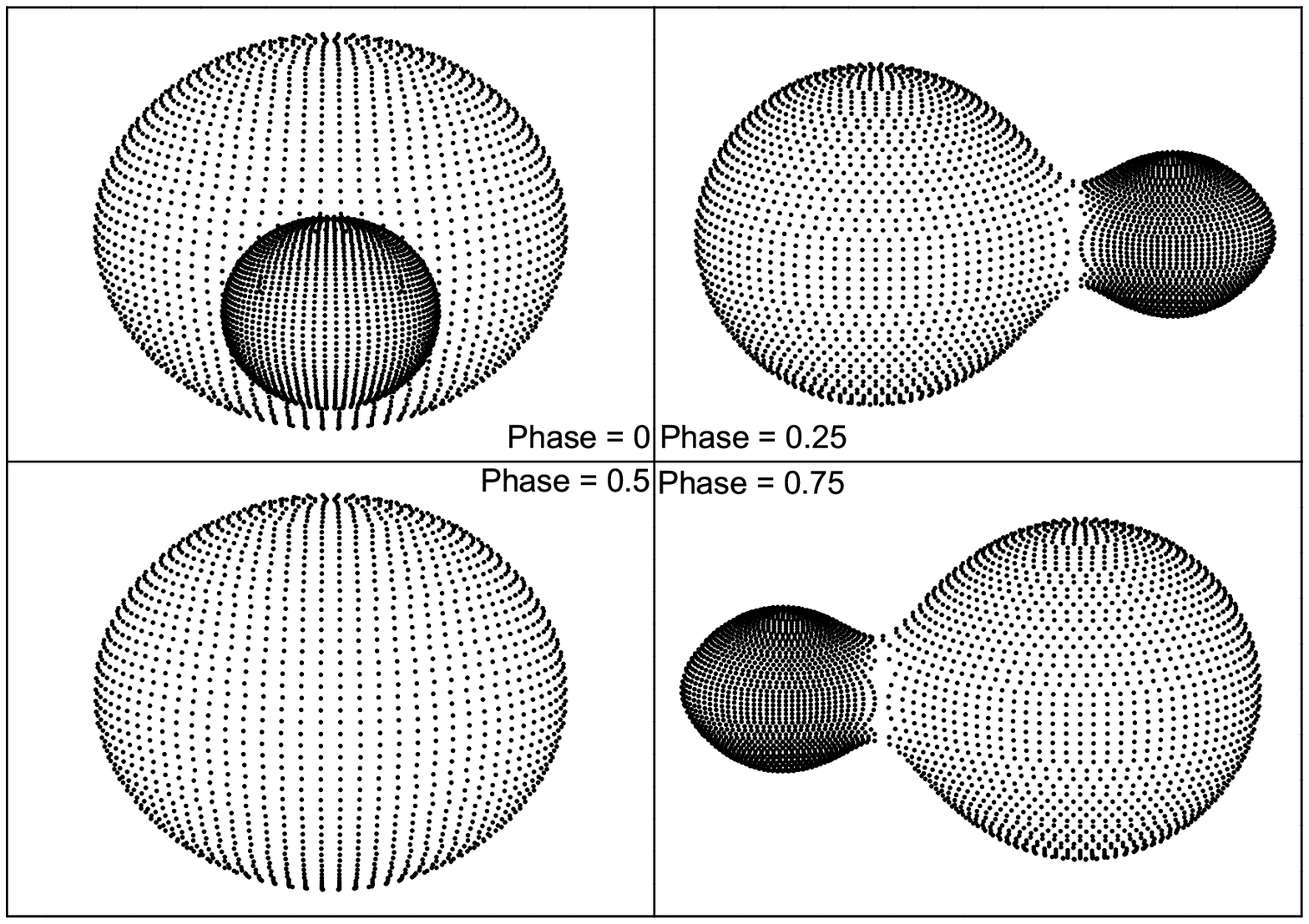}
\caption{Contact configurations of II UMa at phase 0.0, 0.25, 0.5, 0.75.}
\end{center}
\end{figure}

It has to be mentioned that the errors listed in Table 4 are internal errors resulting from the application of the WD code to the supplied data. The effective temperature of the primary star ($T_1$) may have some uncertainties due to the very wide spectral lines coming from the binary and the presence of the lines coming from the third body. We use the color index of II UMa to estimate the real uncertainties of $T_1$. According to the color index (B - V = 0.447) given by the Tycho-2 Catalogue \citep{2000A&A...355L..27H}, the spectral type of II UMa is F5. However, the 2MASS All Sky Catalogue gives the color index of J - H = 0.189 \citep{2003yCat.2246....0C}, which corresponds to a spectral type of F3. And the color index of V - K = 1.089 also supports that the spectral type is F5 \citep{Cox2000}. Thus, the effective temperature of the primary star ($T_1$) may range from 6550K to 6680K. Solution B of $T_1 = 6680K$ is also list in Table 4. According to \citet{2007ASPC..362...82O}'s work, we also set $T_1 = 6412K$ and give Solution C in Table 4. We can conclude that Solution A, Solution B and Solution C give out almost consistent results although $T_1$ ranges from 6412K to 6680K.

\section{Discussions and Conclusions}
Light curves' solutions indicate that II UMa is an overcontact binary system with a quite high contact degree ($ f=86.6\,\%$) and an extremely low mass ratio ($q=0.172$), which indicate that it is at the final evolutionary stage of cool short-period binaries. It may merge into a single rapid-rotation star, which may be the progenitor of blue straggler or FK Com-type star \citep{2015AJ....150...83Z}. The two components have nearly the same surface temperature ($\Delta T = 4K$) in spite of their quite different masses and radii, which indicate that the system is under thermal contact. Considering the mass function given by \citet{2002AJ....124.1738R}: $(M_1+M_2)sin^3i = 2.180\pm0.080M_\odot$ and the orbital inclination ($i = 77.8^{\circ}$) obtained by the light curves' solutions, the mass of the two components are calculated to be $M_1 = 1.99(\pm0.08)M_\odot$, $M_2 = 0.34(\pm0.01)M_\odot$. Spectroscopic search carried out by \citet{2006AJ....132..650D} confirmed that II UMa was a triple system with a solar-type tertiary component, and determined the effective temperature of the third component to be $T_3 = 6100K$, which corresponded to a G0V type star. The mass of the third component is estimated to be $M_3 = 1.34(\pm0.05)M_\odot$ according to the mass ratio ($M_3/M_1 = 0.67$) obtained. The third light ($l_3$) is also included as an adjustable parameter during the photometric processing. The light curve solutions also confirm that it is a triple system and the third component contributes nearly a quarter of the total luminosity. As shown in Fig. 3, the existence of the third light reduces the occultation depth apparently.

Spectroscopic observations show that II UMa may be a contact binary with its primary component to be a giant star. It does have a quite long period (P=0.82522d) which does not obey the well-defined period-color relation of contact binary. However, the parameters obtained by us show that the radius of the apparent giant is equal to about 2.7$R_\odot$. The 2$M_\odot$ star should have a radius equal to about 3.5$R_\odot$ when it leaves MS. Thus, the primary star of II UMa has evolved from Zero Age Main Sequence (ZAMS), but it is still before Terminal Age
Main Sequence (TAMS). II UMa is one of the A-type W UMa stars. Detailed modeling by \citet{2011AcA....61..139S} concluded that initially detached binary systems will eventually evolve to main sequence contact binaries or Algol-type binary systems. We assume that II UMa is formed from an initially detached system via angular momentum loss (AML) by means of magnetic stellar wind. It is just under the late evolutionary stage of late-type tidal-locked binary stars, which might be close to merging and evolving into a single rapid-rotating star.

II UMa is a member of a visual binary system. The close binary system is even confirmed to be a triple system with a close-in solar-type tertiary component orbiting around the close binary system. Thus, it is actually a quadruple system. As discussed by \citet{2013ApJS..209...13Q,2014AJ....148...79Q}, the existence of an additional stellar component in the binary system may play an important role for the formation and evolution by removing angular momentum from the central binary system during the early dynamical interaction or late evolution. The $O$-$C$ curve analysis shows a continuous period increase at a rate of $dP/dt=4.88\times{10^{-7}}day\cdot year^{-1}$, which may be just a part of a cyclic period change, or the combinational period change of a parabolic variation and a cyclic one. More times of minimum light are needed to confirm this. II UMa is an important target for testing theories of star formation and stellar dynamical evolution and interaction. It is possible that third-body interactions in the birth environment may help to accelerate the orbital evolution of the central binary system. Angular momentum is drained from the inner close pair either by the ejection of the tertiary companion \citep{2004A&A...414..633G} or through the Kozai mechanism \citep{1962AJ.....67..591K,2007ApJ...669.1298F}.

\acknowledgments{We thank the anonymous referee for useful comments and suggestions that have improved the quality of the manuscript. This work is supported by the Chinese Natural Science Foundation (Grant No. 11133007 and 11325315), the Strategic Priority Research Program ``The Emergence of Cosmological Structure'' of the Chinese Academy of Sciences (Grant No. XDB09010202) and the Science Foundation of Yunnan Province (Grant No. 2012HC011). New CCD photometric observations of II UMa were obtained with the 60cm and 1m reflecting telescope at Yunnan Observatories (YNOs).}

\end{document}